\documentclass{desyproc}

\usepackage{subfigure}
\usepackage{sidecap}

\newcommand{\prd}{Pys. Rev. D}

\begin{document}
\title{Long-term evolution of massive star explosions}

\author{
{\slshape
T. Fischer$^{1,2}$,
M. Liebend{\"o}rfer$^3$,
F.-K. Thielemann$^3$,
G. Mart{\'i}nez-Pinedo$^{2,1}$,
B.~Ziebarth$^1$
and
K. Langanke$^{1,2,4}$
\\[1ex]}
$^1$ GSI Helmholtzzentrum f\"ur Schwerionenforschung,
   Planckstra{\ss}e~1, 64291 Darmstadt, Germany\\
$^2$ Technische Universit{\"a}t Darmstadt,
  Schlossgartenstra{\ss}e 9, 64289 Darmstadt, Germany\\ 
$^3$ University Basel, Department of Physics, Klingelbergstra{\ss}e 82, 4056 Basel, Switzerland\\
$^4$ Frankfurt Institute for Advanced Studies, Ruth-Moufang
  Stra{\ss}e 1, Frankfurt, Germany
}

\contribID{xy}

\confID{1964}  
\desyproc{DESY-PROC-2010-01}
\acronym{PLHC2010} 
\doi  

\maketitle

\begin{abstract}
We examine simulations of core-collapse supernovae in spherical symmetry.
Our model is based on general relativistic radiation hydrodynamics with
three-flavor Boltzmann neutrino transport.
We discuss the different supernova phases, including the long-term evolution
up to 20~seconds after the onset of explosion during which the neutrino
fluxes and mean energies decrease continuously.
In addition, the spectra of all flavors become increasingly similar, indicating
the change from charged- to neutral-current dominance.
Furthermore, it has been shown recently by several groups independently,
based on sophisticated supernova models, that collective neutrino flavor
oscillations are suppressed during the early mass-accretion dominated
post-bounce evolution.
Here we focus on the possibility of collective flavor flips between electron and
non-electron flavors during the later, on the order of seconds, evolution after the
onset of an explosion with possible application for the nucleosynthesis of heavy
elements.
\end{abstract}

\section{Introduction}

\begin{SCfigure}
\includegraphics[width=0.68\textwidth]{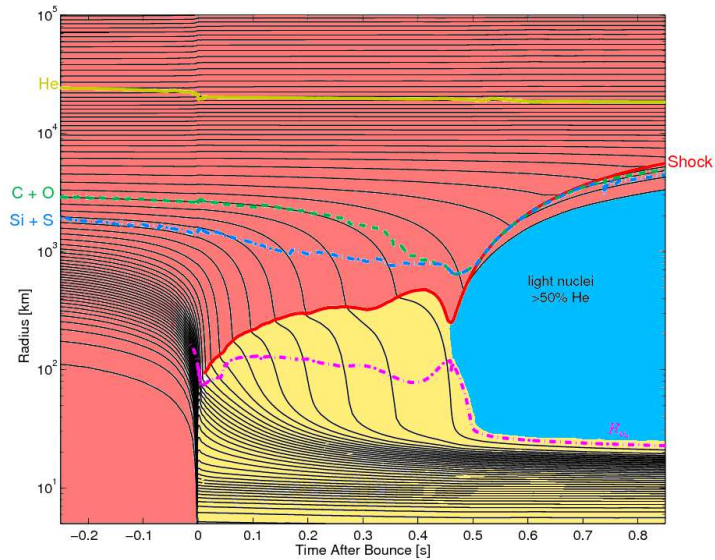}
\caption{Sketching the evolution of selected mass elements during core collapse,
bounce, post-bounce accretion and onset of explosion.
Color coding is according to the dominant composition (light red: heavy nuclei,
blue: $^4$He, yellow: light nuclei and free nucleons).
The solid red and dash-dotted magenta lines mark the positions of shock and
neutrinosphere.
The dashed lines mark the evolution of interfaces between different composition
layers of the progenitor.}
\label{fig:shellplot}
\end{SCfigure}

Explosions of massive stars are related to the formation of a shock wave, which
forms when the collapsing stellar core bounces back at nuclear matter density.
During collapse, the stellar core deleptonizes so that a low central proton-to-baryon
ratio, given by the electron fraction of $Y_e\simeq0.3$, is reached at bounce.
The conditions obtained at bounce depend sensitively on the weak interaction
scheme and the equation of state used.
Fig.~\ref{fig:shellplot} illustrates the radial evolution of selected mass elements.
Before bounce, the infalling mass elements correspond to the central iron-core while
the outer layers of the progenitor are basically unaffected from the central happenings.
After bounce, the shock wave propagates outwards and stalls on a timescale of
5--20~ms due to energy losses from heavy-nuclei dissociation and $\nu_e$-escapes
emitted via large numbers of electron captures during the shock passaged across the
neutrinospheres.
As a result of energy loss, the expanding dynamic bounce shock turns into a standing
accretion shock (SAS).
For the early shock propagation and the position of the $\nu_e$-sphere, see the
red solid and magenta dash-dotted lines in Fig.~\ref{fig:shellplot}.
The post-bounce evolution is given by mass accretion onto the SAS and neutrino
heating, dominantly via $\nu_e$ and $\bar\nu_e$ absorption at the dissociated free
nucleons, behind the SAS on timescales on the order of 100~ms.

Several explosion mechanisms have been explored;
the magneto-rotational~\cite{LeBlanc:1970kg},
the dumping of acoustic energy~\cite{Burrows:2005dv}
and
the standard scenario due to neutrino heating~\cite{Bethe:1985ux}.
Recently, it has been shown that a quark-hadron phase transition can lead to the
formation of an additional shock wave that can trigger
explosions~\cite{Sagert:2008ka,Fischer:2011}.
In this article, we explore standard neutrino-driven explosions in spherical symmetry
of the low-mass 8.8~M$_\odot$ O-Ne-Mg-core and more massive iron-core progenitors.
For the latter, where neutrino-driven explosions cannot be obtained in spherical
symmetry, we enhance neutrino heating in order to trigger explosions.
Fig.~\ref{fig:shellplot} illustrates the standard neutrino-driven explosion of a
15~M$_\odot$ progenitor, for which the accretion phase ends at about 450~ms post
bounce with the onset of explosion.
The SAS turns into a dynamic shock which expands continuously to increasingly
larger radii (see the red solid line in Fig.~\ref{fig:shellplot}).
It has been speculated that collective neutrino flavor oscillations, during the post-bounce
accretion phase, may affect neutrino luminosities and hence heating and cooling.
Recently, it has been shown that matetr dominance suppresses collective flavor
oscillations during the accretion phase~\cite{Chakraborty:2011a,Chakraborty:2011b}.
It has been confirmed by several different groups based on different supernova
models~\cite{Dasgupta:2011,Sarikas:2011}.

At the onset of explosion, mass accretion vanishes and the central proto-neutron star
contracts rapidly (see the magenta dash-dotted line in Fig.~\ref{fig:shellplot}).
It formed at core bounce and is hot and lepton-rich, in which terms it differs
from the final supernova remnant neutron star. 
Between the expanding shock wave and the central proto-neutron star forms a
region of low density and high entropy, where the surface of the proto-neutron star
is subject to continued neutrino heating.
There, a low-mass outflow develops known as {\em neutrino-driven wind}.
The first sophisticated radiation-hydrodynamics study of the neutrino-driven
wind was a milestone of research in the field~\cite{Woosley:1994ux}.
It could explain the solar $r$-process abundances, due to the obtained strong
wind with high entropies per baryon $\sim300$~k$_B$ and generally neutron-rich
conditions with $Y_e\simeq$0.35--0.48.
The neutrino-driven wind has also long been explored in static steady-state
models~\cite{Duncan:1986,Hoffman:1996aj,Thompson:2000tu,Thompson:2001ys}
and dynamic studies~\cite{Woosley:1992,Woosley:1994ux,Takahashi:1994yz,Witti:1994},
as possible site for the nucleosynthesis of heavy
elements~\cite{Otsuki:1999kb,Wanajo:2006mq,Wanajo:2006ec}.
However, recent supernova simulations that include Boltzmann neutrino transport cannot
confirm the early results.
They obtain generally proton-rich conditions and entropies per baryon on the order
of 100~k$_B$~\cite{Fischer:2009af,Huedepohl:2010}.
The main difference to the early studies is related to the evolution of neutrino
luminosities and mean energies.
Within the current models, they reduce continuously during the proto-neutron star
deleptonization on timescales on the order of 10~seconds after the onset of explosion.
Furthermore, the $\nu_e$ and $\bar\nu_e$ spectra become increasingly similar.
Charged-current dominace reduces, because final state electrons become
Pauli-blocked and nucleons become degenerate at the neutrinospheres,
due to the increasing density.
Instead, the spectra become dominated by neutral-current processes during the
proto-neutron star deleptonization.

The relevance of collective neutrino flavor oscillations has long been investigated in various astrophysical applications. Although collective neutrino flavor oscillations are suppressed during the accretion phase, they may be relevant after the onset of explosion due to the continuously decreasing matter density in the presence of still high neutrino densities.Here, we explore the possibility of complete spectral flips of $\bar\nu_e$ and $\bar\nu_{\mu/\tau}$ at a fixed flip energy and their impact to $\nu p$-process nucleosynthesis for a selected trajectory from a supernova simulation of a massive iron-core progenitor. We find that it enhances the neutron production rate which in turn increases the production of heavy nuclei with $A>90$.

The manuscript is organized as follows. We will summarize main aspects of our core-collapse model in \S~2. In \S~3 we will illustrate standard neutrino-driven explosions of massive stars in spherical symmetry as well as the neutrino-driven wind phase after the onset of an explosion. \S~4 is devoted to the evolution of neutrino luminosities and spectra. Illustration of our simplified neutrino flavor flip analysis and the impact to nucleosynthesis will be discussed in \S~5. We close with a summary in \S~6.

\section{Core-collapse supernova model}

The simulations under investigation are based on general relativistic radiation
hydrodynamics and three-flavor Boltzmann neutrino transport in spherical symmetry.
For details, see the following references~\cite{Mezzacappa:1993gn,
Mezzacappa:1993gm,Mezzacappa:1993gx,
Liebendoerfer:2001a, Liebendoerfer:2001b,Liebendoerfer:2002,Liebendoerfer:2004}.
Recent improvements of the adaptive mesh have been added in
ref.~\cite{Fischer:2009af}.
It enables large stable timesteps and allows for long simulation times on the order
of 10~seconds.
The list of weak processes considered is given in Table~\ref{table-nu-reactions},
including references.
In addition, the implementation of the following weak process,
$\nu_e + \bar\nu_e \leftrightarrows \nu_{\mu/\tau} + \bar\nu_{\mu/\tau}$,
has been discussed in ref~\cite{Fischer:2009}, following ref.~\cite{Buras:2002wt}.
For the current study, weak magnetism corrections as well as $N$--$N$--recoil and
ion-ion-correlations have not been included in the weak processes.

\begin{table}[htp!]
\centering
\begin{tabular}{ccc}
\hline
\hline
&
weak process$^1$
&
References
\\
\hline
1
&
$\nu_e + n \leftrightarrows p + e^-$
&
\cite{Bruenn:1985en}\\
2
&
$\bar{\nu}_e + p \leftrightarrows n + e^+$
&
\cite{Bruenn:1985en}
\\
3
&
$\nu_e + (A,Z-1) \leftrightarrows (A,Z) + e^-$
&
\cite{Bruenn:1985en}
\\
4
&
$\nu + N \leftrightarrows \nu' + N$
&
\cite{Bruenn:1985en}
\\
5
&
$\nu + (A,Z) \leftrightarrows \nu' + (A,Z)$
&
\cite{Bruenn:1985en}
\\
6
&
$\nu + e^\pm \leftrightarrows \nu' + e^\pm$
&
\cite{Bruenn:1985en,Mezzacappa:1993gx}
\\
7
&
$\nu + \bar{\nu} \leftrightarrows e^- + e^+$
&
\cite{Bruenn:1985en,Mezzacappa:1993gx}
\\
8
&
$\nu + \bar{\nu} + N + N \leftrightarrows N + N$
&
\cite{Hannestad:1997gc}
\\
\hline
\hline
&
$^1$
Notes: $\nu=\{\nu_e,\bar{\nu}_e,\nu_{\mu/\tau},\bar{\nu}_{\mu/\tau}\}$, $N=\{n,p\}$
&
\end{tabular}
\caption{Neutrino reactions considered, including references.}
\label{table-nu-reactions}
\end{table}

For matter in nuclear statistical equilibrium (NSE), the equation of state from
ref.~\cite{Shen:1998gg} was used.
It is based on relativistic mean field approach and the Thomas-Fermi approximation for
heavy nuclei, with a simplified composition of neutrons, protons, $\alpha$-particles and
a single representative heavy nucleus with average atomic mass $A$ and charge $Z$.
Baryon contributions for matter in non-NSE are added using a slim nuclear reaction
network for 20 nuclei (see~\cite{Fischer:2009af} and references therein).
It is used only for energy production.
On top of the baryons, contributions from ($e^-$, $e^+$) and photons as well as
ion-ion-correlations for non-NSE are added~\cite{Timmes:1999}.

Our core-collapse simulations are launched from the low-mass 8.8~M$_\odot$
O-Ne-Mg-core~\cite{Nomoto:1984,Nomoto:1987} and from more massive iron-cores
of 10.8, 15 and 18~M$_\odot$~\cite{Woosley:2002zz}.
Their evolution during accretion and explosion, as well as the long-term evolution
on timescales on the order of seconds after the onset of explosion, will be discussed
in the next section.

\begin{figure}[htb!]
\centering
\includegraphics[width=0.99\textwidth]{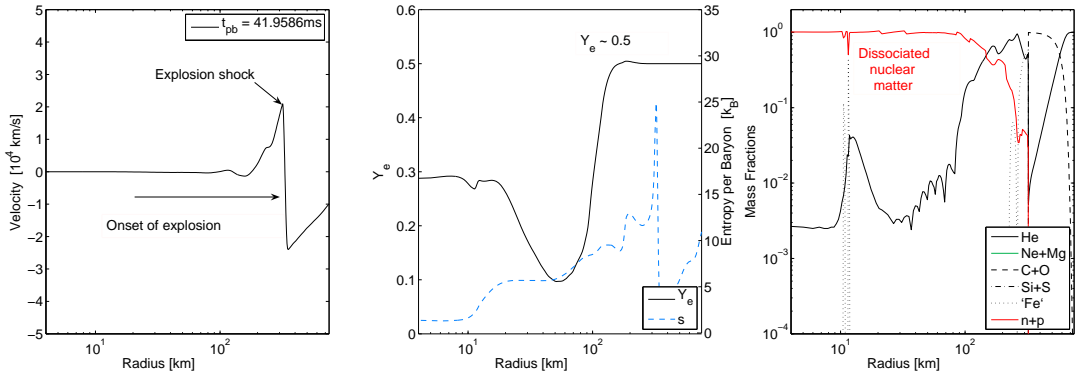}
\caption{Radial profiles of selected quantities at the onset of explosion, for the 8.8~M$_\odot$ O-Ne-Mg-core progenitor (data are taken form ref.~\cite{Fischer:2009af}).}
\label{fig:expl.onset-n08c}
\end{figure}

\section{Explosions and long-term evolution}

Neutrino-driven explosions of the low-mass 8.8~M$_\odot$ O-Ne-Mg-core can be obtained even in spherically symmetric supernova models~\cite{Kitaura:2006,Fischer:2009af}. The success of this model is related to the special structure of the progenitor. Only about 0.1~M$_\odot$ of the 1.376~M$_\odot$ core is composed of iron-group nuclei, at the onset of collapse. The outer layers are dominated by $^{20}$Ne and $^{24}$Mg as well as further out $^{12}$C and $^{16}$O. During collapse, the Ne and Mg layers are partly burned to iron-group elements and hence the enclosed mass inside the iron-core grows. Moreover, when the standing accretion shock reaches the interface between C-O and He-layers, where the density drops over more than 10 orders of magnitude, it turns into a dynamic shock with positive velocities. It determines the onset of explosion, at about 35~ms post bounce, after which the shock expands continuously to larger radii. Fig.~\ref{fig:expl.onset-n08c} illustrates the onset of explosion for this model (see ref.~\cite{Fischer:2009af} and references therein). During the early explosion phase the $\bar\nu_e$and $\nu_e$-spectra are very similar. Note further, for this low-mass progenitor the shock wave expands basically into vacuum due to the extremely low density of the He-rich hydrogen envelop, where velocities on the order of the speed of light are reached. The competition between reactions (1) and (2) in Table~\ref{table-nu-reactions}, lead to even slightly neutron-rich conditions with $Y_e\simeq0.4681$--0.4986 during the initial shock expansion after the onset of explosion between about 200--400~ms post bounce. The timescale for $\bar\nu_e$ captures to turn material to the proton-rich side is not sufficient. In axially symmetric simulations, matter becomes even more neutron-rich early after the onset of explosion, developing mushroom-like pockets with $Y_{e,\text{ min}}\simeq 0.404$~\cite{Wanajo:2011a}. It may be a possible site for the weak $r$-process, producing elements with atomic mass between $A>$56--90, for which the production problem based on standard chemical evolution models has been discovered in ref.~\cite{Qian:2001}. Recently, the question of an additional nucleosynthesis process required in order to explain the observed abundances of these elements has been addressed. It became known as light-element primary process (LEPP)~\cite{Travaglio:2004} and is an active subject of research.

\begin{figure}[htb!]
\centering
\includegraphics[width=0.99\textwidth]{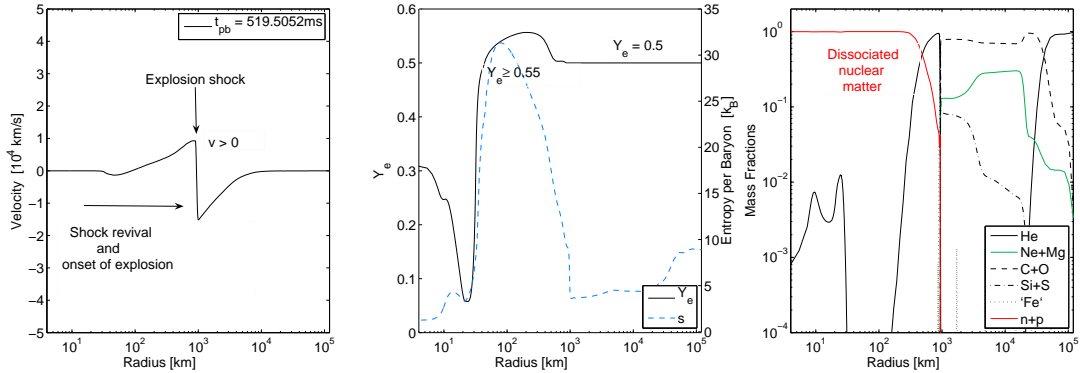}
\caption{Radial profiles of selected quantities at the onset of explosion, for the 15~M$_\odot$ iron-core progenitor.}
\label{fig:expl.onset-h15l}
\end{figure}

The situation is different for more massive stars, illustrated in Fig.~\ref{fig:expl.onset-h15l} at the example of a 15~M$_\odot$ iron-core progenitor. The extended high-density Si-S-layer surrounding the more massive iron-core, leads to a post-bounce accretion phase that can last for several 100~ms (depending on the progenitor model). The central proto-neutron star is much more compact at the onset of explosion. The luminosities of $\nu_e$ and $\bar\nu_e$ are very similar during the post-bounce accretion phase as well as at the onset of explosion. Hence, due to the rest-mass difference between neutrons and protons, matter becomes proton-rich with $Y_e\simeq$0.5--0.57. The magnitude of the differences between $\nu_e$ and $\bar\nu_e$ luminosities and mean energies, and consequently $Y_e$, is an active subject of research. It may change taking corrections from weak magnetism and improved weak rates into account. Note that the explosions for the iron-core progenitors under investigation are obtained applying enhanced heating and cooling rates (detailed balance is fulfilled), in order to trigger the explosions. For more details, see ref.~\cite{Fischer:2009af}. Core-collapse simulations based on multi-dimensional models, that include sophisticated neutrino transport~\cite{Ott:2007,Marek:2009,Bruenn:2010ae}, are required for simulation times on the order of several seconds after the onset of explosion.

\begin{figure}[htb!]
\centering
\subfigure[Radial profiles of $Y_e$ and velocity.]{
\includegraphics[width=0.45\textwidth]{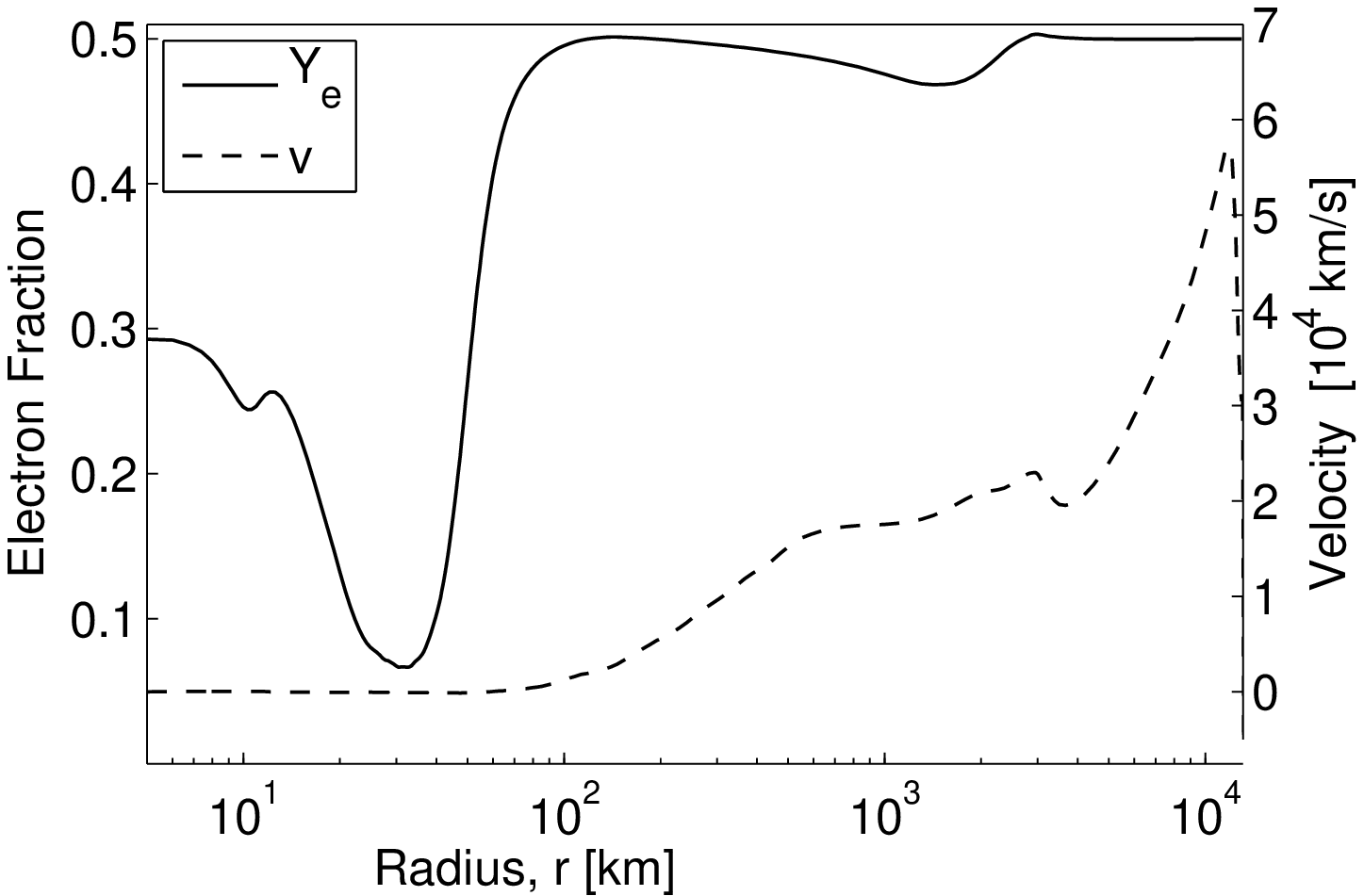}
\label{fig:ye-n08c}}
\subfigure[Heating rate during the neutrino-driven wind]{
\includegraphics[width=0.50\textwidth]{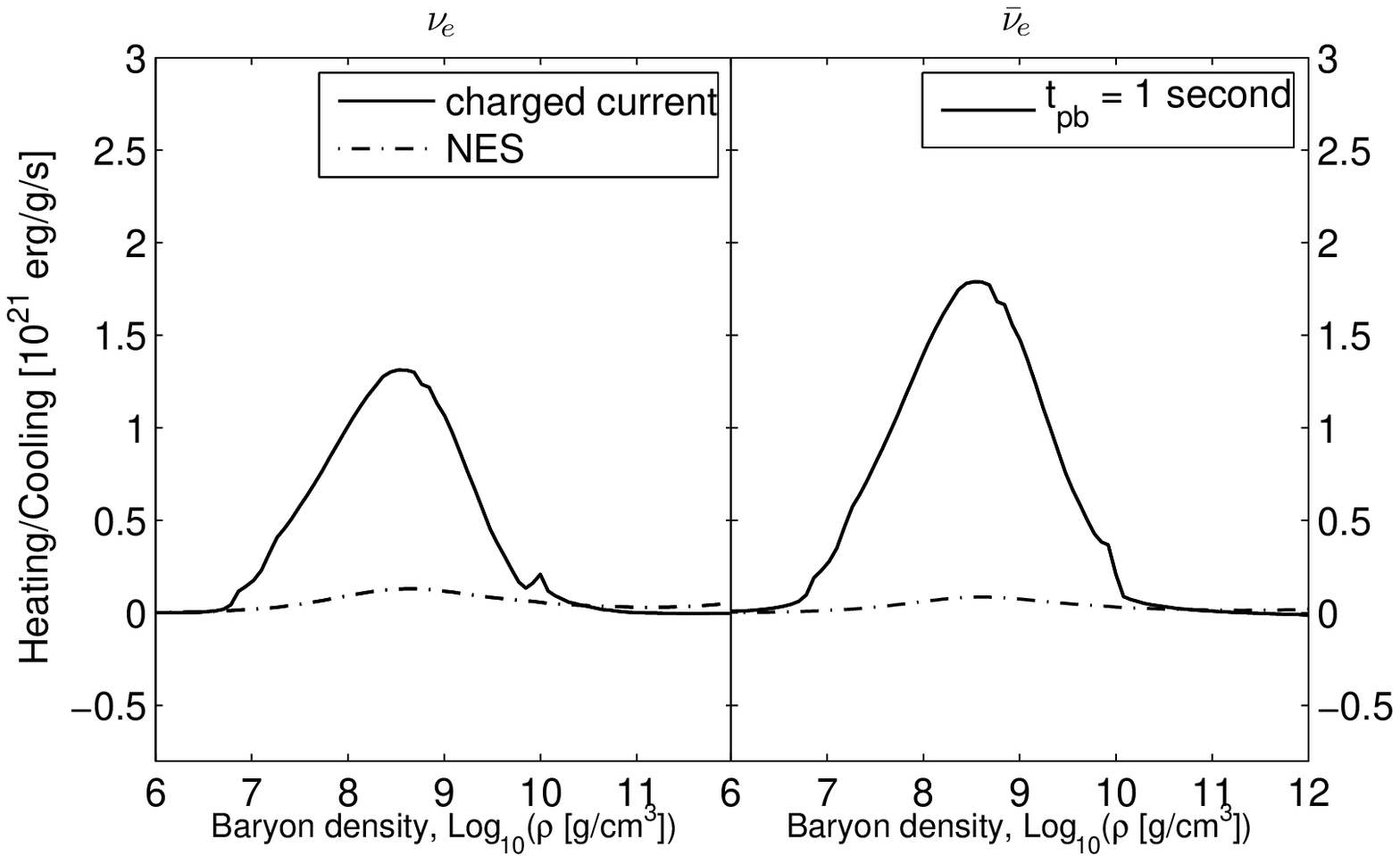}
\label{fig:heatplot-h10a}}
\caption{Electron fraction and velocity profiles at the early explosion phase at about 200~ms post bounce for the 8.8~M$_\odot$ model in graph~\ref{fig:ye-n08c} and net energy-deposition rates at the onset of the neutrino-driven wind phase at about 1~second post bounce in graph~\ref{fig:heatplot-h10a} (data are taken form ref.~\cite{Fischer:2009af}).}
\end{figure}

\begin{figure}[hb]
\centering
\subfigure[Baryon density and Temperature]{
\includegraphics[width=0.48\textwidth]{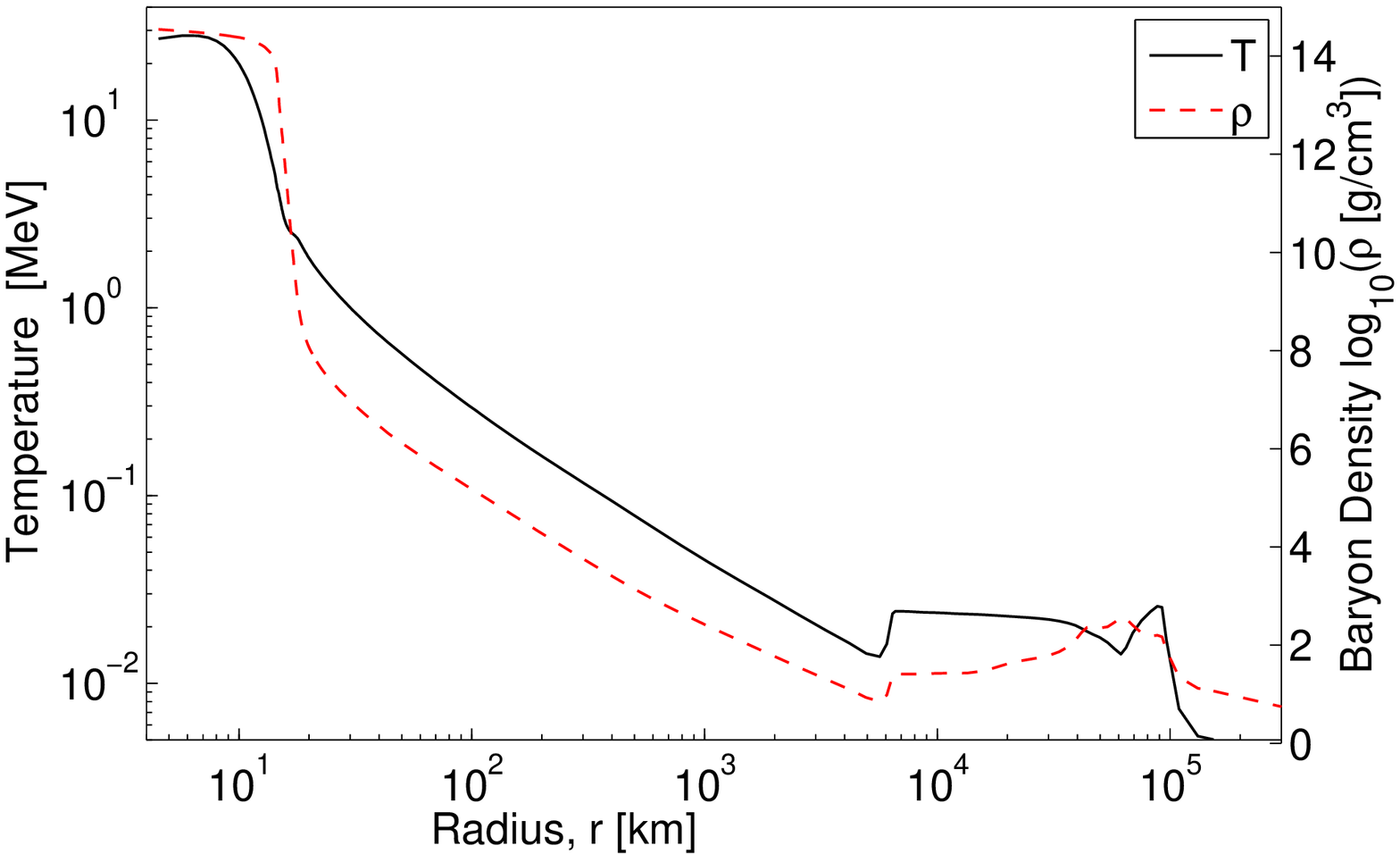}
\label{fig:density-h10a}}
\subfigure[Velocity and entropy per baryon.]{
\includegraphics[width=0.48\textwidth]{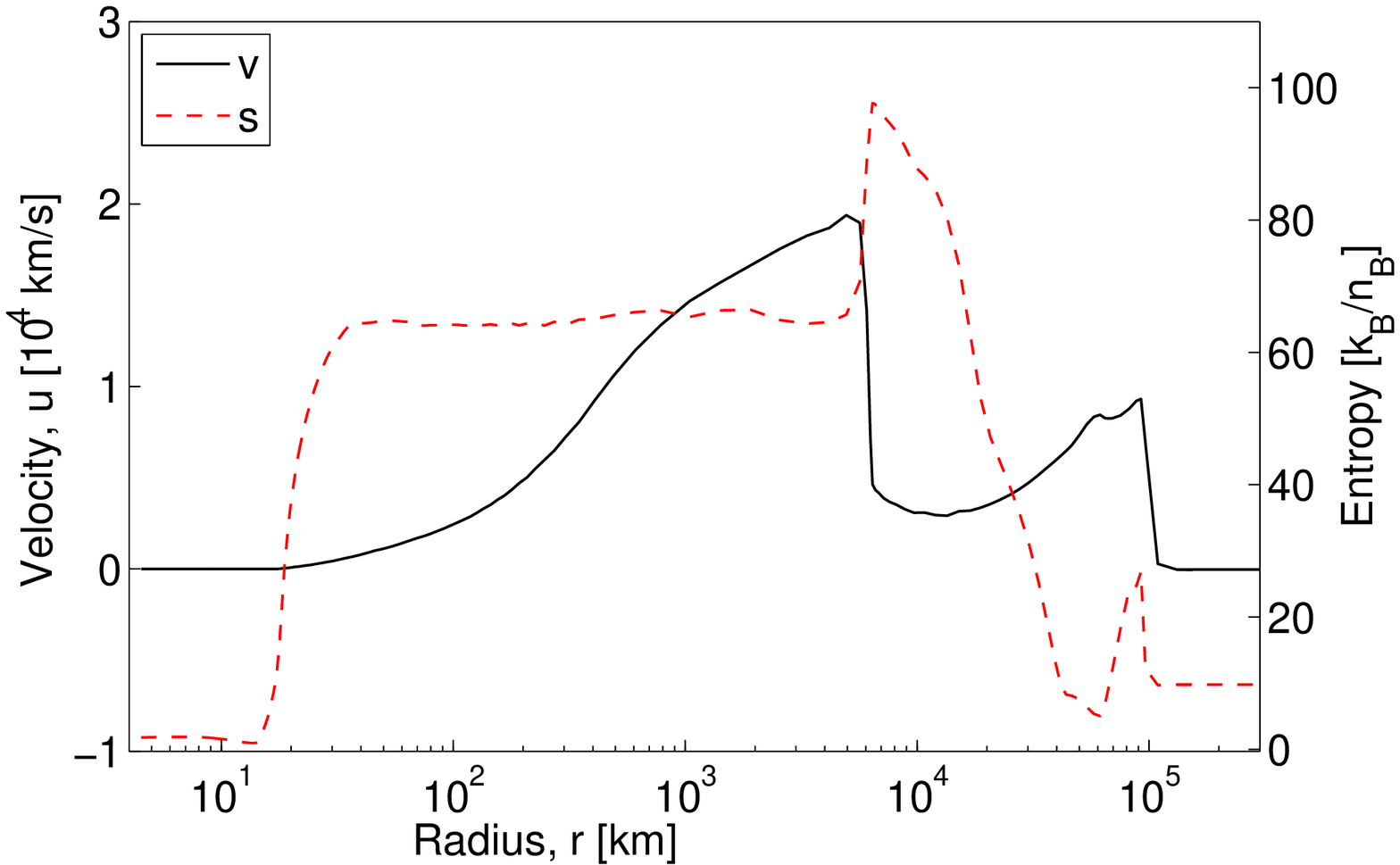}
\label{fig:velocity-h10a}}
\caption{Radial profiles of selected quantities during the neutrino-driven wind phase for the 10.8~M$_\odot$ progenitor model under investigation (data are taken form ref.~\cite{Fischer:2009af}).}
\label{fig:wind}
\end{figure}

After the onset of explosion a region of low density and high entropy develops between the expanding explosion shock and the central proto-neutron star. Moreover, at the surface of the proto-neutron star establishes net-heating, illustrated in Fig.~\ref{fig:heatplot-h10a}. It leads to a low-mass outflow, known as the neutrino-driven wind.
%
%
%
Compared to the very fast initial expansion of the 8.8~M$_\odot$ model, the situation is different for more-massive iron-core progenitors. There, the neutrino-driven ejecta expand into the extended C-O and He-layers with baryon densities between $10^1$~g~cm$^{-3}$ to $10^3$~g~cm$^{-3}$, where also the shock expansion slows down. For illustration, see the radial profiles at a selected post-bounce time during the neutrino-driven wind phase in Fig.~\ref{fig:density-h10a} at the example of the 10.8~M$_\odot$ progenitor. Moreover, the neutrino-driven wind collides with the slower moving explosion shock. Note that in case of a super-sonic neutrino-driven wind, a reverse shock forms as shown in Fig.~\ref{fig:velocity-h10a}. This leads to an additional temperature and entropy increase (see Figs.~\ref{fig:density-h10a} and \ref{fig:velocity-h10a} between 5000-6000~km). The impact of the reverse shock on possible nucleosynthesis has been investigated recently~\cite{Arcones:2011a}. At late times, neutrino heating reduces and the neutrino-driven wind turns back to sub-sonic velocities before it vanishes completely.

\section{Neutrino spectra evolution}

\begin{SCfigure}
\centering
\includegraphics[width=0.635\textwidth]{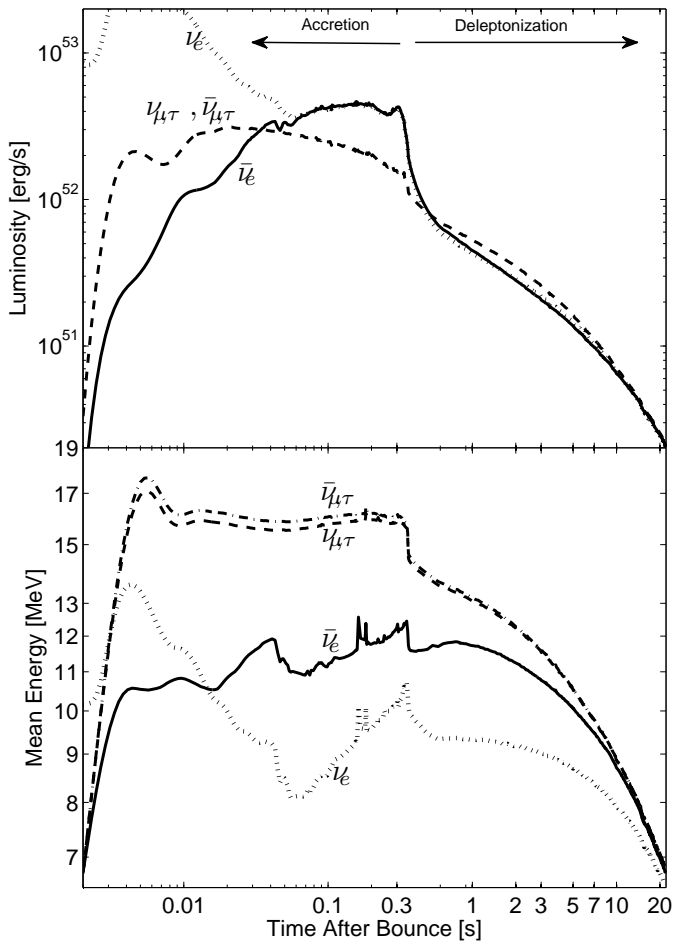}
\caption{Post bounce evolution of neutrino (dotted lines: $\nu_e$, solid lines:
$\bar\nu_e$, dashed lines: $\nu_{\mu/\tau}$, dash-dotted lines: $\bar\nu_{\mu/\tau}$)
luminosities (top) and mean energies  (bottom) for the 18~M$_\odot$ progenitor model
(data are taken form ref.~\cite{Fischer:2009af}).
At the end of the accretion phase, indicated by the sharp jumps at about 350~ms
post bounce, the luminosities of all flavors decrease continuously.
The same holds for the mean energies.
Furthermore, neutrino luminosities and spectra become increasingly similar for all
flavors during the deleptonization phase.
They converge for this model at about 20~seconds post bounce.
It is related to the reducing dominance of charged-current reactions during the
proto-neutron star deleptonization.
Instead, the spectra are dominated by neutral-current reactions (neutrino-neutron scattering).}
\label{fig:lumin-h18b}
\end{SCfigure}

The evolution of neutrino luminosities and mean energies is shown in 
Fig.~\ref{fig:lumin-h18b} at the example of the 18~M$_\odot$ progenitor
up to 22~seconds post bounce.
The observables are sampled in the co-moving reference frame at
a distance of 500~km, well outside the neutrinospheres.

The $\nu_e$-luminosity, $\mathcal{O}(10^{52})$~erg/s, rises slowly after the
deleptonization burst has been launched at 20~ms post bounce.
$\bar\nu_e$ and $\nu_{\mu/\tau}$ are produced only after bounce.
The $\nu_{\mu/\tau}$-luminosities rises until about 20~ms post bounce and the
$\bar\nu_e$-luminosity rises continuously.
After reaching their maximum, the $\nu_{\mu/\tau}$-luminosity decreases slowly
continued during the accretion phase on timescales of 100~ms.
Furthermore, the $\nu_e$ and $\bar\nu_e$ luminosities are determined by mass
accretion at the neutrinospheres.
They rise slowly on a timescale of 100~ms and reach their maximum of several
$10^{52}$~erg/s at the onset of explosion at about 350~ms post bounce.
After that, mass accretion vanishes and the electron flavor luminosities decrease
rapidly one order of magnitude within the first second after the onset of explosion.
The $\nu_{\mu/\tau}$-luminosity reduces accordingly and takes similar values as
the electron flavor luminosities.
The magnitude of the differences between the different flavors is an active
subject of research.
It depends sensitively on the weak processes considered and the dimensionality
of the model.
On a long timescale on the order of several 10~seconds, i.e. the proto-neutron star
deleptonization, the neutrino luminosities of all flavors reduce below $10^{50}$~erg/s.
Furthermore, they become practically indistinguishable.
In multi-dimensional models, and in the presence of aspherical explosions,
mass accretion is still possible after the onset of an explosion.
A possible enhancement of the neutrino fluxes remains to be shown for
simulation times on the order of several seconds.

The mean energies, shown at the bottom of Fig.~\ref{fig:lumin-h18b}, have a similar
behavior as the neutrino fluxes.
They rise during the early post bounce phase up to 12, 14 and 19~MeV for
$\nu_e$, $\bar\nu_e$ and $\nu_{\mu/\tau}$ respectively, at the onset of explosion.
After that, $\langle E \rangle_{\nu_{\mu/\tau}}$ decreases continuously.
$\langle E \rangle_{\nu_{e}}$ and $\langle E \rangle_{\bar\nu_{e}}$ stay about
constant until about 2~seconds post bounce, after which they decrease as well.
The spectra of all flavors converge during the evolution after the onset of explosion,
i.e. the difference between the mean energies of all flavors reduces continuously during
the proto-neutron star deleptonization.
For the 18~M$_\odot$ progenitor model under investigation, the spectra have
converged at about 20~seconds post bounce.
It is related to the reduced dominance of charged-current reactions, due to final-state
electron blocking and increasing nucleon degeneracy.
Instead, the spectra are dominated by neutral-current reactions, in particular
neutrino-neutron scattering.

The small, and even reducing, difference between $\nu_e$ and $\bar\nu_e$
luminosities and spectra has important consequences for the composition.
It leads to generally proton-rich conditions with $Y_e\simeq$0.52--0.56 for matter
that becomes gravitationally unbound during the proto-neutron star deleptonization
in the neutrino-driven wind.
The results obtained for the low-mass O-Ne-Mg-core are in qualitative agreement
with the results of the Garching group~\cite{Huedepohl:2010}.
The magnitude of $Y_e$ obtained for the models under investigation depends
sensitively on the equation of state and the weak processes used.

\section{Nucleosynthesis under proton-rich conditions}

Matter at the surface of the proto-neutron star is in NSE, due to the high temperatures and densities. During the expansion in the neutrino-driven wind, matter cools until reaching larger distance from the proto-neutron star surface, where nucleons recombine into heavy nuclei. This nucleosynthesis depends sensitively on the initial proton-to-baryon ratio. It is determined via the competition of reactions (1) and (2) from Table~\ref{table-nu-reactions} in the dissociated regime at the proto-neutron star surface. It depends on the above discussed electron flavor luminosities and spectra. In the presence of similar $\nu_e$ and $\bar\nu_e$ luminosities and mean energies\footnote{For neutron-rich conditions, $ \varepsilon_{\bar\nu_e} -  \varepsilon_{\bar\nu_e} \lesssim 4\Delta$, where $\varepsilon = \langle E^2\rangle/\langle E \rangle$. $\langle E \rangle$ is the mean neutrino energy and $\langle E^2 \rangle$ is the square value of the root-mean-square (rms) energy and the neutron-proton rest-mass difference $\Delta =1.2935$~MeV}, matter becomes proton-rich due to the neutron-proton rest-mass difference~\cite{Qian:1996}. The proton-rich conditions obtained lead to isospin symmetric nuclei, mainly $^{56}$Ni, as well as $^4$He and free protons. However, the further nucleosynthesis stops at, e.g., $^{64}$Ge which has a long beta-decay half-life of $\simeq64$~s (known as waiting-point nucleus) and because $^{65}$As has a low proton separation energy of $\simeq90$~keV.

\begin{figure}[htp!]
\centering
\subfigure[]{
\includegraphics[width=0.48\textwidth]{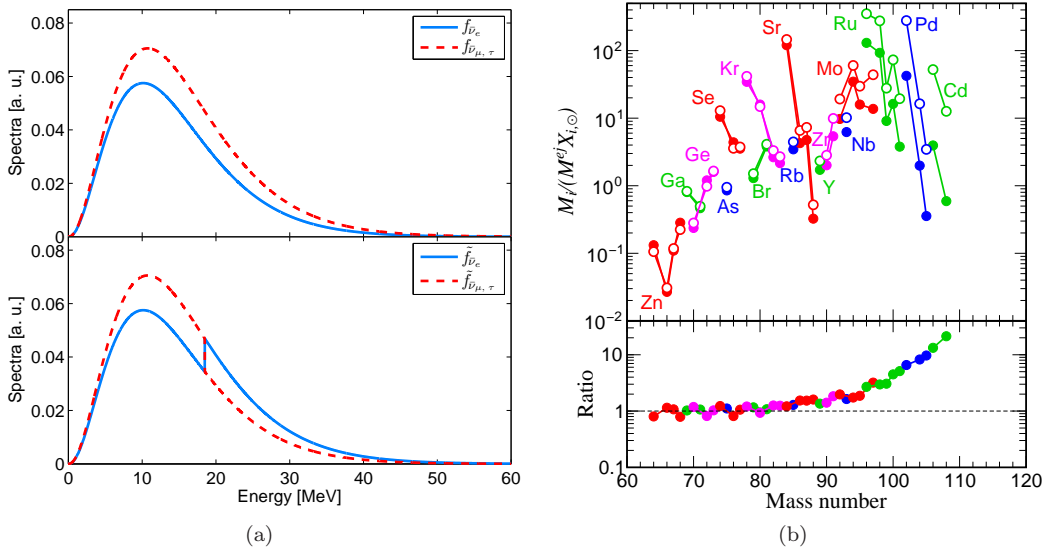}
\label{fig:spectra}}
\subfigure[]{
\includegraphics[width=0.45\textwidth]{abund-14.eps}
\label{fig:abund}}
\caption{Anti-neutrino spectra with (bottom) and without (top) spectral split in graph~\ref{fig:spectra} and overproduction factors based on $\nu p$-process nucleosynthesis in graph~\ref{fig:abund} (both figures are taken from  ref.~\cite{MartinezPinedo:2011}).}
\end{figure}

The situation changes including neutrino reactions, mostly $\bar\nu_e$ because isospin symmetric nuclei are inert to $\nu_e$ captures. $\bar\nu_e$ can be captured at mainly protons as well as nuclei on timescales of seconds at distances of several 100~km when temperatures are as low as several $10^9$~K. With the consequently increased neutron density, waiting-point nuclei can be overcome via $(n,p)$-reactions, and the mass flow can continue to heavier nuclei. This process is know as $\nu p$ process~\cite{Frohlich:2005ys, Pruet:2006,Wanajo:2006ec}. Which heaviest nuclei can be reached depends on the conditions obtained in the neutrino-driven wind. The full circles in the upper panel of Fig.~\ref{fig:abund} illustrate the final abundances of a particular mass element from a simulation of a 15~M$_\odot$ progenitor star~\cite{Buras:2005rp,Pruet:2005,Pruet:2006}, labelled ''1116~ms''. The figure shows the ratio $M_i/(M^\text{ej}X_{i,\odot})$, where $M_i$ is the produced mass of isotope $i$ with the corresponding solar mass fraction $X_{i,\odot}$ and total mass ejected (taken form ref.~\cite{Pruet:2005}). The neutrino spectra required are taken from ref.~\cite{Pruet:2006} and approximated by an $\alpha$-fit (for details, see~\cite{MartinezPinedo:2011} and references therein).

Collective neutrino flavor oscillations have long been investigated in the context of core-collapse supernovae. Here, we explore possible effects of complete collective flavor flips between $\bar\nu_e$ and $\bar\nu_{\mu/\tau}$ on the $\nu p$ process, assuming normal mass hierarchy. Following ref.~\cite{Dasgupta:2009}, Fig.~\ref{fig:spectra} illustrates the spectral flip taking place at energy of about 18~MeV (bottom) in comparison to the unmodified spectra (top). Note that including nucleon recoil, especially for heavy-lepton neutrinos, will decrease high-energy spectral differences between $\bar\nu_e$ and $\bar\nu_{\mu/\tau}$ and hence the enhanced high-energy tail of the oscillated $\bar\nu_e$ as illustrated in Fig.~\ref{fig:spectra} will be reduced.

For the $\nu p$ process, $\bar\nu_e$ spectra are required. In addition to the unmodified neutrino spectra, we include the oscillated spectra into the nucleosynthesis analysis and repeat the calculation. Relevant is the change of the neutron production rate due to the inclusion of the spectral flip. It increases the neutron production due to the enhanced high-energy tail of the flipped $\bar\nu_e$ spectra. Furthermore, we parametrized the flip energy~\cite{MartinezPinedo:2011}. We found that the enhancement of the neutron production rate is a robust result which is independent from the flip energy. It depends on the characteristics of the neutrino spectra obtained in particular simulations. Result of the nucleosynthesis outcome, including the complete spectral flip, is shown in Fig.~\ref{fig:abund} (open circles). The neutron-production rate is enhanced by a factor of 1.4, according to the spectra taken from ref~\cite{Pruet:2005}. It significantly increases the production of heavy elements with $A>90$. Furthermore, overproduction factors for nuclei with $A=64$, 68 and 76 are reduced slightly. It has been discussed in more details in ref.~\cite{MartinezPinedo:2011}.

\section{Summary}

Core-collapse supernova simulations are investigated in spherical symmetry.
We focused on the post-bounce accretion phase and the evolution after the onset
of explosion, illustrating different conditions comparing the two intrinsically different
core-collapse progenitors with an O-Ne-Mg-core and an iron-core.
The explosion of the first one is a combination of neutrino heating and energy
deposition from nuclear burning, on a short timescale of only few 10~ms post bounce.
Furthermore, matter remains slightly neutron-rich for this progenitor only during the
early explosion phase.
On the other hand, the massive Si-layer surrounding the iron-core for more massive
progenitors leads to an extended post-bounce accretion phase.
It can last several 100~ms (depending on the progenitor model) during which
the central proto-neutron star, which formed at core bounce, contracts continuously.
The resulting similar electron flavor neutrino luminosities lead to generally
proton-rich conditions at the onset of explosion.

After an explosion has been launched, continued neutrino heating at the proto-neutron
star surface leads to a low-mass outflow on a timescale on the order of seconds.
It became known as neutrino-driven wind, for which we discussed and illustrated
typical conditions.
Furthermore, the neutrino luminosities and spectra of all flavors become
increasingly similar during the long-term proto-neutron star deleptonization.
It is a consequence of the decreasing importance of charged-current reactions,
because at the neutrinospheres (a) electrons become Pauli-blocked, (b) the number of
neutrinos reduces continuously and (c) nucleons become degenerate.
Instead, the spectra are dominated by scattering at neutrons.
The reducing difference between $\nu_e$ and $\bar\nu_e$ leads to generally
proton-rich conditions.
It has important consequences for the nucleosynthesis of heavy elements and
may allow for the $\nu p$ process.

Recently, it has been shown that neutrino oscillations are dominated by matter terms
and hence collective flavor oscillations are suppressed during the accretion
phase~\cite{Chakraborty:2011a,Chakraborty:2011b,Dasgupta:2011,Sarikas:2011}.
However, they may be possible during the later evolution after the onset of an
explosion, during which the matter density decreases continuously.
In addition to the standard nucleosynthesis in proton-rich conditions,
we explore the possibility of the $\nu p$ process assuming a complete spectral
flip between $\bar\nu_e$ and $\bar\nu_{\mu/\tau}$ at a certain split energy.
We used neutrino spectra from recent supernova simulations.
It results in enhanced $\bar\nu_e$ captures, dueÄ to the enhanced high-energy tail of
$\bar\nu_e$, which in turn increases the number of neutrons present during
$\nu p$-process nucleosynthesis.
It allows the matter flow to proceed to heavier nuclei and results in larger
abundances of nuclei with $A>64$.
The current analysis has to be improved, using full energy- and angle-dependent
neutrino oscillation techniques, in order to compute the spectral evolution consistently
in massive star explosions.

\section*{Acknowledgement}

The work was supported by the Swiss National Science Foundation under project
numbers~PBBSP2-133378, PP00P2-124879/1 and 200020-122287 as well as HIC for
FAIR and the Helmholtz Alliance EMMI and the SFB 634 at the Technical University
Darmstadt.
Parts of the work has been initiated during the GSI Summer School Program.
B.Z. thanks the program for financial support.
The authors are additionally supported by CompStar, a research networking
program of the European Science Foundation.
%

\begin{footnotesize}


\bibliographystyle{unsrt}

\end{footnotesize}


\end{document}